# Chemical modeling of exoplanet atmospheres

O. Venot[1,*] and M. Agúndez[2,3]

[1] Instituut voor Sterrenkunde, Katholieke Universiteit Leuven, Celestijnenlaan 200D, 3001 Leuven, Belgium
[2] Univ. Bordeaux, LAB, UMR 5804, F-33270, Floirac, France
[3] CNRS, LAB, UMR 5804, F-33270, Floirac, France

**Abstract**

The past twenty years have revealed the diversity of planets that exist in the Universe. It turned out that most of exoplanets are different from the planets of our Solar System and thus, everything about them needs to be explored. Thanks to current observational technologies, we are able to determine some information about the atmospheric composition, the thermal structure and the dynamics of these exoplanets, but many questions remain still unanswered. To improve our knowledge about exoplanetary systems, more accurate observations are needed and that is why the Exoplanet Characterisation Observatory (EChO) is an essential space mission. Thanks to its large spectral coverage and high spectral resolution, EChO will provide exoplanetary spectra with an unprecedented accuracy, allowing to improve our understanding of exoplanets. In this work, we review what has been done to date concerning the chemical modeling of exoplanet atmospheres and what are the main characteristics of warm exoplanet atmospheres, which are one of the main targets of EChO. Finally we will present the ongoing developments that are necessary for the chemical modeling of exoplanet atmospheres.

## 1. Introduction

Determining the elemental composition of the atmosphere/envelope of exoplanets is one of the keys to understand their origin and evolution. In the Solar System, elemental abundances in the envelope of the gaseous giants and icy giants are not well constrained because of the condensation cold trap that removes some carriers of the main elements (in particular oxygen) from the upper atmosphere probed by observations. Interestingly, it could be easier to unveil the elemental composition of an extrasolar hot/warm- Jupiter or Neptune than that of our own Jupiter and Neptune. This, however, requires the ability to translate spectroscopic observations into molecular abundances and physical conditions (pressure and temperature) and then into an elemental composition. This second step would be relatively simple if atmospheres were at chemical equilibrium, in which case the detailed molecular composition depends solely on the temperature, pressure, and elemental composition. Except for the very hottest exoplanets, equilibrium does not hold due to UV photolysis, mixing, and advection. As a consequence, the kinetics of hundreds of individual reactions controls the atmospheric composition.

The diversity of exoplanetary atmospheres observable with EChO spans a broad range of physical conditions. Individual reaction rates must therefore be known at temperatures ranging from below room temperature to above 2500 K and - because the deep atmospheric layers are chemically mixed with the layers probed by spectroscopic observations – at pressures up to about 100 bars. Fortunately, these conditions are also found in typical combustion experiments and explosion engines and we benefit from decades of research on the chemical kinetics needed for exoplanetary atmospheres. Through a collaboration with research departments dedicated to applied combustion, we were able to develop a kinetic model based on a network of ~2000 of reactions that has been experimentally validated over this temperature-pressure range and for gaseous mixtures of 105 species made of H, He, C, O and N. This kinetic model is now implemented into atmosphere models that can accurately simulate the chemistry of any type of atmosphere that are dominated by these elements.

## 2. Recent progress in 1D atmospheric chemical modeling

The first 1D models used to study the atmosphere of transiting exoplanets assumed thermodynamic equilibrium because of the high temperature of these atmospheres (e.g. Burrows and Sharp, 1999; Seager and Sasselov, 2000; Sharp and Burrows, 2007; Barman, 2007; Burrows et al., 2007, 2008; Fortney et al., 2008a). However, photodissociations by the stellar UV fluxes, and transport/mixing resulting from the dynamics powered by the intense irradiation, occur on timescales comparable or even shorter than the chemical timescales. As a consequence, planetary atmospheres are out of equilibrium and hot/warm exoplanets are the site of a unique chemistry where thermochemical kinetics (involving exothermic and endothermic chemical reactions) competes with photolysis and gas dynamics. Departures from equilibrium have been first addressed based on tions mescales arguments (e.g. Lodders and Fegley, 2002; Fortney et al., 2006, 2008b; Visscher et al., 2006, 2010; Madhusudhan and Seager, 2010) or modeled by coupling the dynamical processes with some chemical reactions describing the $CO$-$CH_4$ conversion (Cooper and Showman, 2006).
Robust and versatile modeling of atmospheric chemistry requires the use of a kinetic model, including all relevant individual chemical and photochemical processes.

* Email : olivia.venot@ster.kuleuven.be

The first (photo)chemical models for exoplanets were adapted from Jupiter chemical schemes that did not include endothermic reactions, occurring at high temperature (Liang et al. 2003, 2004). These reactions can be neglected in Jupiter's atmosphere, but are crucial to model the atmosphere of most known transiting exoplanets. A next generation of models then included these endothermic reactions (Zahnle et al. 2009a,b; Line et al. 2010; Moses et al. 2011) consistently with the constraints of thermodynamics. These kinetics models evolve towards the **thermo**chemical equilibrium when photolysis and transport are turned off (The effect of these out of equilibrium processes will be seen in Sect. 3). Different networks with very different individual kinetic rates can, however, evolve toward the same equilibrium provided that they use the same equilibrium constant to derive the rates of endothermic reactions from exothermic ones. In a next step, kinetic networks and their associated rates have therefore been constructed and selected by their ability to reproduce the global chemical behavior of experimentally-controled abundances. This has been made possible thanks to a close collaboration with teams working on applied combustion who deal with mixtures whose elemental composition (H, O, C, N), temperature (300-2500 K), and pressures (up to hundreds of bars) overlap with those of exoplanet atmospheres (see for instance Bounaceur et al. 2007). The accuracy demanded when modeling the chemistry of an explosion engine exceeds by far those needed to predict the atmospheric composition of exoplanets. This experimental validation confers a great confidence on the results obtained over a very broad range of conditions. The chemical scheme by Venot et al. (2012) includes all the reactions required to model the kinetic evolution of radicals and molecules containing fewer than three carbon atoms. It involves 105 compounds, linked by ~2000 reactions. It is available to the scientific community on the database KIDA (Wakelam et al. 2012, http://kida.obs.u-bordeaux1.fr).

## 3. Applications to hot Jupiters and warm Neptunes

Chemical models of exoplanetary atmospheres have largely focused on the study of the hot Jupiters HD 209458b and HD189733b, as these two objects have been widely observed at primary transit and secondary eclipse conditions, leading to important constraints on their atmospheric chemical composition (Zahnle et al. 2009a,b; Line et al. 2010; Moses et al. 2011; Venot et al. 2012). The degree of departure from chemical equilibrium is different in the atmosphere of these two exoplanets, as illustrated in Figure 1. In the case of HD 189733b, one can clearly notice the larger effect of photolyses and vertical mixing on the chemical composition, with all species affected, except the main reservoirs ($H_2$, $H_2O$, $CO$, and $N_2$), showing important departures from chemical equilibrium.

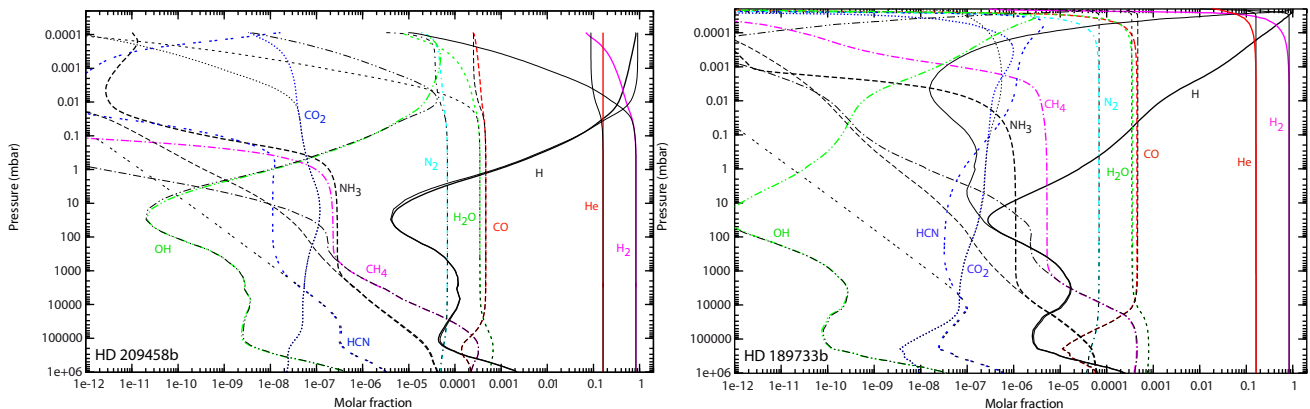

Figure 1: Steady-state composition of HD 209458b (*left*) and HD 189733b (*right*) calculated with a non equilibrium model (color lines), compared to the thermodynamic equilibrium (thin black lines). See more details in Venot et al. (2012).

The atmospheres of hot and warm Neptunes are also very interesting sites to be studied with chemical models (Line et al. 2010; Moses et al. 2013, Agúndez et al. 2014a, Venot et al. 2014). The warm Neptunes GJ 3470b and GJ436b have an equilibrium temperature lower than hot Jupiters and their atmospheres are more sensitive to disequilibrium processes. Using these two planets as study cases, we have investigated the effect of vertical mixing and UV flux, but also the metallicity of the envelope (which can be much higher for these planetary masses) and the thermal profile. These four parameters have an important influence on the chemical composition. For instance, as illustrated on Figure 2, the abundances of the main reservoirs of carbon, CO and $CH_4$, depend to a large extent on the metallicity and the temperature.

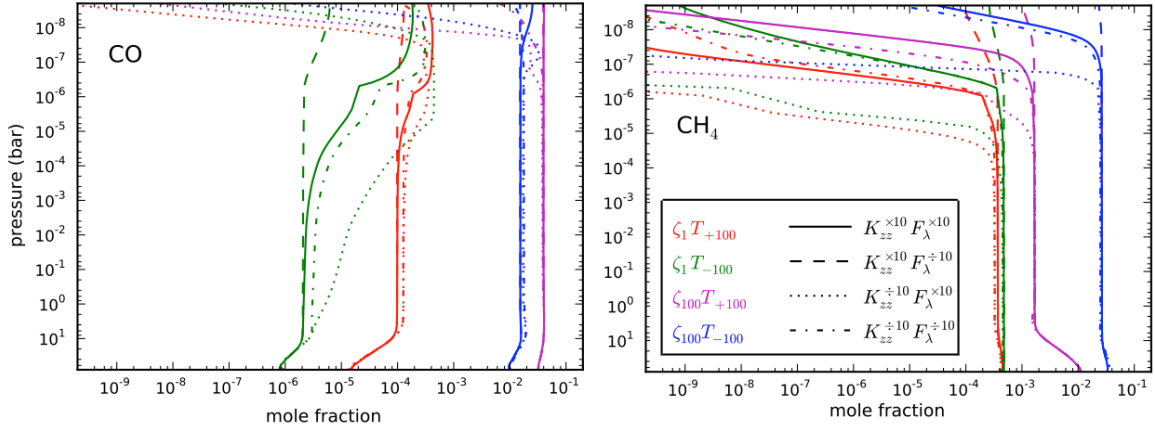

Figure 2: Vertical distribution of the abundances of CO (*left*) and $CH_4$ (*right*) as calculated through 16 models of GJ 3470b in which the space of parameters of metallicity ($\xi$), temperature ($T$), eddy diffusion coefficient ($K_{zz}$), and stellar UV flux ($F_\lambda$) are explored. Each colour corresponds to a set of metallicity and temperature, and each line style to a set of eddy diffusion coefficient and stellar UV flux. See more details in Venot et al. (2014).

## 4. Beyond one-dimensional models

Hot Jupiters and Neptunes orbit close to their host star and are thus expected to be tidally locked, showing permanent day and night sides (Guillot et al. 1996). The unequal heating of the planet maintains a strong temperature contrast between the day and night sides. These temperature gradients and the fact that photochemistry is switched on in the irradiated dayside and off in the non-irradiated nightside might drive important variations in the chemical composition of the atmosphere with longitude and latitude.

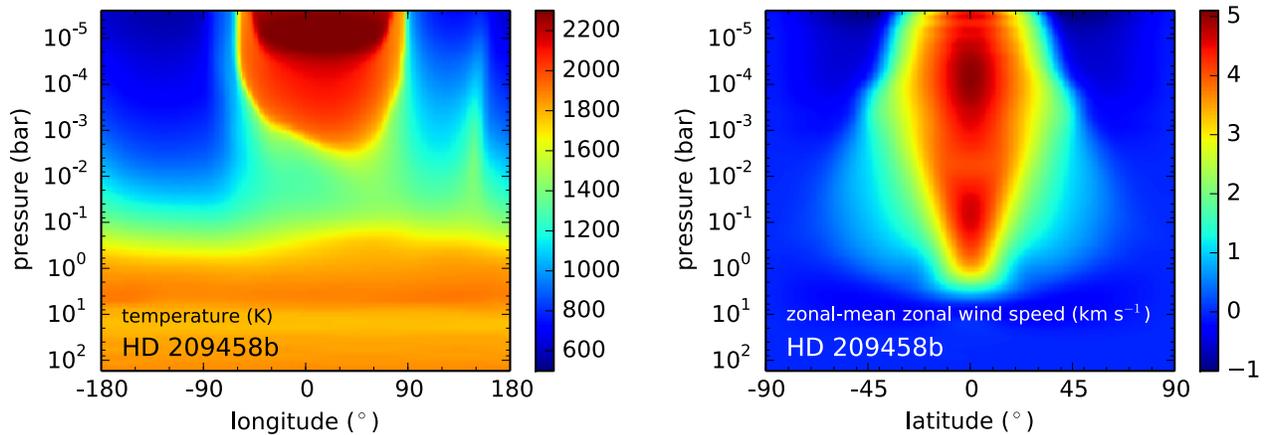

Figure 3: *Left*: Temperature structure averaged latitudinally over ±20° around the equator of HD 209458b. Note the existence of strong temperature contrasts between the day and night sides, with an extremely hot dayside stratosphere above the 1 mbar pressure level. *Right*: Zonal-mean zonal wind speed (positive is eastward and negative westward) as a function of latitude and pressure. Note the existence of a superrotating wind above the 1 bar pressure level in the equatorial region (±20° in latitude). Both are calculated with a GCM simulation of HD 209458b (Parmentier et al. 2013). See more details in Agúndez et al. (2014b).

These inhomogeneities can be alleviated, or even almost suppressed, by the strong horizontal winds present in the atmospheres of these planets, which tend to redistribute the heat from the day to the night side and to homogenize the chemical composition among the different regions of the atmosphere. Evidences of strong winds, with velocities of a few km s$^{-1}$, in the atmosphere of hot Jupiters and Neptunes come from both general circulation models (GCMs), which predict that circulation is dominated by a strong equatorial superrotating jet (Cho et al. 2008; Showman et al. 2009; Lewis et al. 2010 ; Dobbs-Dixon et al. 2012; Rauscher & Menou 2013; Parmentier et al. 2013) and observations of phase curves showing an offset between the hot spot and the substellar point (Knutson et al. 2007) and Doppler shifts in spectral lines attributed to day-to-night winds (Snellen et al. 2010). The existence of winds and horizontal gradients in the temperature and chemical composition of hot Jupiter atmospheres has mainly been considered from a theoretical point of view, although some of their effects can be studied through phase curve observations (Fortney et al. 2006; Cowan & Agol 2008; Majeau et al. 2012; de Wit et al. 2012), monitoring the transit ingress and egress (Fortney et al. 2010), and Doppler shifts of spectral lines during the primary transit (Snellen et al. 2010; Miller-Ricci Kempton & Rauscher 2012; Showman et al. 2013). It is important to note that even if observations are not fine enough to probe 2D or 3D effects in great detail, the study of 2D and 3D atmospheric processes is important as they may determine to a great extent the global average properties of the atmosphere.

Therefore, on the one hand, the uneven irradiation of these planets tend to produce an atmosphere with a non homogeneous distribution of temperature and chemical composition, and on the other, horizontal winds tend to homogenize temperature and molecular abundances among the different regions of the planet. Whether the atmosphere shows an irregular or rather uniform structure is the result of these two competing factors. The subject can be studied theoretically and constrained with observations of the emission spectra of exoplanets at different phases and transmission spectra acquired at the transit ingress and egress, a type of observations which are especially well suited for EChO.

The 3D temperature structure of hot atmospheres has been investigated theoretically to a great detail by numerous groups, mainly in the context of general circulation models. An illustration of the important variations of temperature as a function of longitude in the atmosphere of the hot Jupiter HD 209458b is given in Figure 3, as well as an illustration of the variation of the zonal-mean zonal wind speed as a function of latitude. For a detailed description of how such inhomogeneities can be probed by EChO we refer to Parmentier et al. (2014).

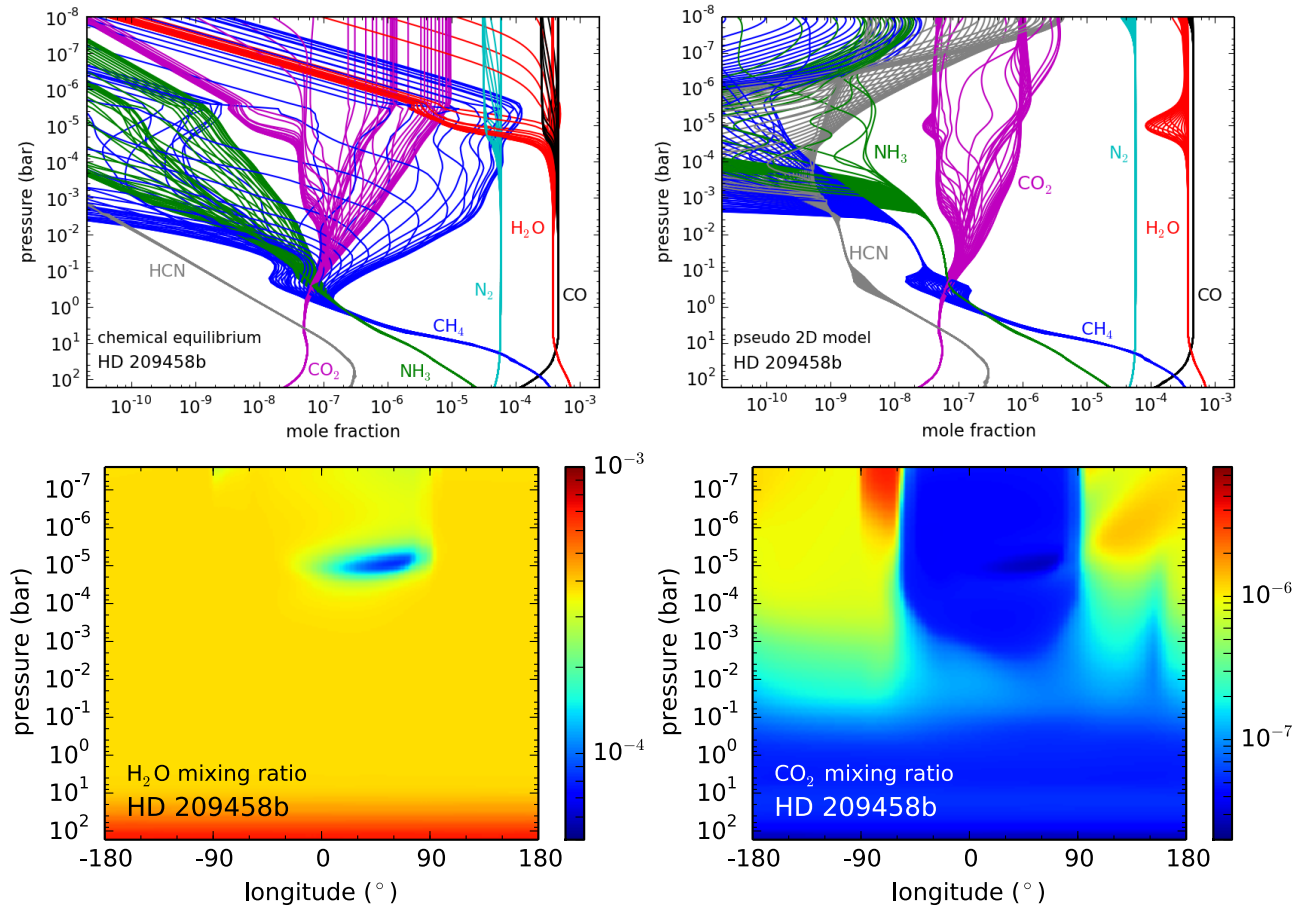

Figure 4: *Up panel*: Vertical cuts of the abundance distributions of some of the most abundant molecules at longitudes spanning the 0-360° range, as calculated with chemical equilibrium (*left*) and with a pseudo two-dimensional chemical model (*right*) for HD 209458b's atmosphere. Note that the large longitudinal variations of the abundances predicted by chemical equilibrium are attenuated to an important extent in the more realistic pseudo two-dimensional model. *Down panel*: Distribution of $H_2O$ (*left*) and $CO_2$ (*right*) as a function of longitude and pressure in the equatorial band of HD 209458b's atmosphere, as calculated with the pseudo two-dimensional chemical model. See more details in Agúndez et al. (2014b).

Here we focus on how atmospheric constituents are distributed in the atmospheres of hot Jupiters and Neptunes according to theoretical calculations. It is clear that to investigate the subject, one-dimensional models need to go one step further toward two- and three-dimensional models to properly deal with the longitude and/or latitude dimensions. Significant contributions to the subject have been made in recent years. A first attempt to understand the interplay between dynamics and chemistry was undertaken by Cooper & Showman (2006), who coupled a three-dimensional general circulation model of HD 209458b to a simple chemical kinetics scheme dealing with the interconversion between CO and $CH_4$. These authors found that, even in the presence of important temperature gradients, the mixing ratios of CO and $CH_4$ are homogenized throughout the planet's atmosphere in the 1 bar to 1 mbar pressure regime. A different approach was adopted by Agúndez et al. (2012), who coupled a robust chemical kinetics scheme to a simplified dynamical model of HD 209458b's atmosphere, in which the atmosphere rotates as a solid body, mimicking a uniform zonal wind. This exploratory study neglected vertical mixing and photochemistry but was able to identify the main effect of zonal winds of the distribution of molecules, which we may summarize as "the zonal wind tends to homogenize the chemical composition, bringing molecular abundances at the limb and nightside regions close to chemical equilibrium values characteristic of the dayside".

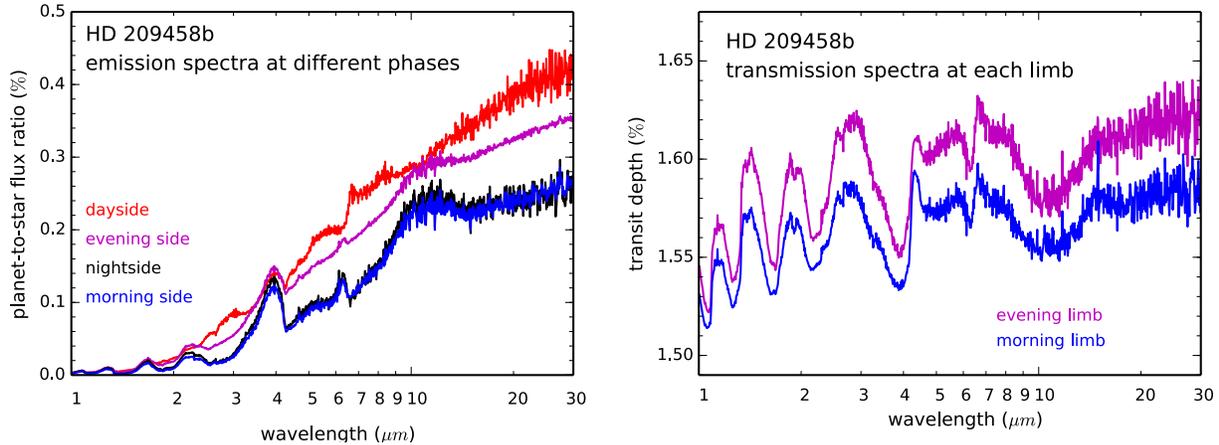

Figure 5: *Left*: Calculated emission spectra for HD 209458b at 4 different phases in which the planet faces the observer the day and night sides, and the two sides in between, centered at the evening and morning limbs. The planetary flux is shown relative to that of the star. *Right*: Calculated transmission spectra for the evening and morning limbs of HD 209458b. The absolute scale of the transmission spectrum is set by assigning the value of the planetary radius to the 1 bar pressure level. Since no attempt has been made to reproduce the absolute scale indicated by primary transit observations, our transit depth values are somewhat higher than given by observations. All spectra have been smoothed to a resolving power R = 300. See more details in Agúndez et al. (2014b).

A step further has been recently taken by Agúndez et al. (2014b) with a pseudo two-dimensional model of the atmospheres of HD 209458b and HD 189733b, which takes into account simultaneously thermochemical kinetics, photochemistry, as well as vertical and horizontal transport, the latter modeled as a uniform zonal eastward wind. The distribution of the chemical composition with altitude and longitude in the atmosphere of these two hot Jupiters is found to be quite complex because of the interplay of the various physical and chemical processes at work. In the main, much of the distribution of molecules is driven by the strong zonal wind and the limited extent of vertical transport, resulting in an important homogenization of the chemical composition with longitude. In general, molecular abundances are quenched horizontally to chemical equilibrium values of the hottest dayside regions, and thus the composition in the cooler nightside regions is highly contaminated by that of warmer dayside regions. The abundance of some molecules such as CO and $H_2O$ show little variation with either longitude or altitude, while others such as $CO_2$ experience significant abundance variations with longitude (see Figure 4). In general, the cooler the planet the largest the homogenization of the chemical composition with longitude. Moreover, in cooler planets such as transiting Neptunes orbiting M dwarfs, the temperature contrast between the day and night sides is less severe because the cooling rate scales with the cube of temperature, and therefore the composition is expected to be even more homogeneous with longitude than in warmer planets such as HD 209458b and HD 189733b. The important homogenization of the chemical composition predicted for hot Jupiter and Neptunes possesses important constraints on how the planetary emission spectrum varies with phase and on the differences between the transmission spectrum of the planet probing each of the two meridians of the planet terminator during the transit ingress and egress (Fig. 5). Most such variations in the spectra are expected to be due to differences in the temperature structure rather than to differences in the chemical composition between different regions of the planetary atmosphere, and this can be readily tested with the observations that EChO will be able to carry out.

## 5.  Ongoing developments

**More elements, more species**

Although they have much smaller cosmic abundances, elements like Sulfur, Phosphorus or Chlorine can play a role in the chemistry and the radiative budget of atmospheres and could be included. Indeed, it has been shown by Zahnle et al. 2009a than Sulfur species can be quite abundant in hot Jupiters atmospheres and absorb in the UV domain. By the way of complex interaction between species, S-species influence the global photochemistry of the atmosphere and could change the abundances of major compounds, like CO or CH4 for instance. Thus, it is important to include these species in the chemical scheme. These species are less relevant for applied combustion and therefore less data are currently available but we are currently working on this question.

**Self-consistent coupling between the 1D thermal structure and the composition**

The chemical composition depends on the temperature in the atmosphere but the temperature depends on the molecular abundances of species that absorb the incident stellar radiation and/or the planetary thermal emission. Iterations between radiative-convective calculations and chemical models are thus required. This has already been tested in 1D (Agúndez et al., 2014a) and is possible provided enough computing time is available. This study shows that variation of temperature due to the use of non-equilibrium, instead of chemical equilibrium, abundances are small (<100 K). This is because because $H_2O$, the main species determining the thermal structure, remains at an abundance close to chemical equilibrium. The extent of these variations depends on the metallicity and becomes more important as this latter increases.

**Phase changes, clouds & aerosols**

There is a growing evidence of cloud signatures in exoplanet spectroscopic observations (Pont et al., 2013) and there are theoretical reasons to believe that condensation should occur. For instance, VO and TiO are expected to be exchanged between the gaseous and solid phase with important consequences on the thermal profile (Spiegel et al. 2009). Phase changes are therefore being implemented into the kinetic models. These phase changes can lead to the formation of aerosols that can affect the chemical composition of exoplanet's atmospheres and the planetary spectra (with scattering and absorption). The effect of Sulfur aerosols has been studied already (Hu et al. 2012, 2013), but further studies need to be done. It is important to characterize the nature of aerosols present in atmospheres in order to interpret the observations of exoplanets correctly.

**Coupling 3D hydrocodes and kinetics**

Current GCM can only include chemical kinetic models based on a very small network of reactions and species. One possible way to deal with this issue is to derive small chemical networks that provide a reasonably good description of the kinetics of a reduced group of species for a specific range of physical conditions and chemical background. For instance, for an atmosphere consisting of Solar abundances and with temperatures in the 800-1400K range, it may be possible to produce a network of less than 20 budget reactions with modified rate coefficients that is able to describe the behaviour of the few species important for the radiative transfer in the atmosphere ($H_2O$, CO, $CH_4$, $CO_2$). It is our objective to produce, test and release such reduced networks.

**High temperature UV cross sections**

Photodissociations by UV play an important role in atmospheric chemistry. Available cross sections are derived from ambient or low temperature measurements. Data corresponding to the high temperatures found in hot Jupiter atmospheres (around 1000 K) are very rare. When some data exist, they have been usually done for combustion studies, on monochromatic beams or restricted wavelength ranges, usually above 190 nm while the 100-200 nm range is extremely important. It is known (e.g Koshi et al., 1991; Schulz et al., 2002; Oehlschlaeger et al., 2004) that the absorption of the UV flux is supposed to increase at wavelengths above 160 nm together with the temperature. So, when using room temperature data to model hot atmospheres we underestimate the absorption of the UV flux and photolysis rates. Consequently, it is crucial to have data corresponding to the temperature of warm exoplanets. In the context of EChO, this lack of data is now addressed by measuring absorption cross section and their temperature dependency for the important molecules of planetary atmospheres in the VUV wavelength range (115 – 230 nm). Thanks to several campaign of measurements at the synchrotron BESSY (Berliner Elektronenspeicherring-Gesellschaft für Synchrotronstrahlung) and at the LISA (Laboratoire Interuniversitaire des Systèmes Atmosphériques), the absorption cross section of $CO_2$ (Venot et al., 2013) has been determined up to 800 K. It can be seen on Fig. 6 that the absorption cross section of $CO_2$ varies significantly, increasing by several orders of magnitude together with the temperature.
The impact of the temperature dependency of the $CO_2$ cross section on exoplanet atmospheres has been studied by Venot et al. (2013) on a prototype planet whose characteristics are similar to those of the hot Neptune GJ 436b.

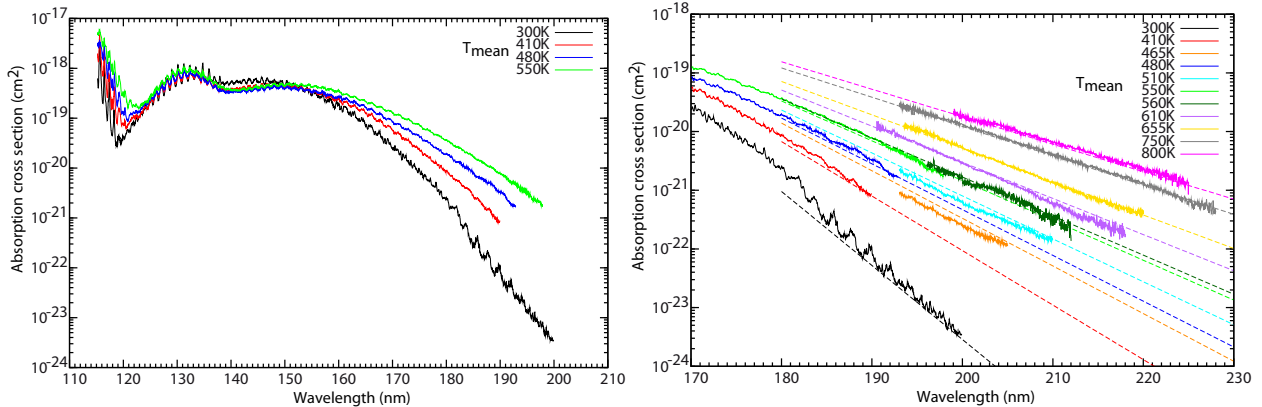

Figure 6: *Left*: Absorption cross section of $CO_2$ at T=300 K (black), 410 K (red), 480 K (green) and 550 K (blue) for wavelengths between 115 and 200 nm. *Right:* Absorption cross section of $CO_2$ for wavelengths longer than 195 nm at 465 K, 510 K, 560 K, 610 K, 655 K, 750 K, and 800 K, plotted with the cross section at ambient temperature (black). The absorption cross sections calculated with the parameterization are plotted with the same color coding.

**Influence of the star**

The host star of the planet can show some variability, i.e. flares, that could have an impact on the atmospheric composition of the planet. For instance, in case of a sudden high energy release, the photodissociation rates could be increased, which would have for consequent to destroy/create species. Thus, the planetary spectrum would be modified. For the interpretation of this spectrum, it is crucial to quantify the effect of such stellar variability.

## 6. Conclusion: chemical modeling and EChO

Robust chemical modeling is a key requirement for EChO or any instrument aiming at understanding the nature of exoplanetary atmospheres. Chemical modeling is required to derive the atmospheric elemental compositions from spectroscopic observations. For giant planet envelope, this elemental composition can be described in terms of metallicity and C/O but for planets of a more terrestrial nature, whose atmospheres are outgassed from accreted solids, a much broader range of compositions is expected (Forget & Leconte 2013). Chemical models do also provide the link between composition and spectral features but also a way to explore the diversity of exoplanet atmospheres and associated signatures before the mission and to prepare the field of research to this diversity. It is indirectly required to interpret observations (for instance orbital spectrophotometry) in terms of circulation/dynamics. Indeed, the chemical longitudinal gradients can influence the shape of the lightcurve and the detailed composition determines the pressure of the layers probed by observations. An atmosphere with the same 3D thermal structure can exhibit either strong or negligible phasecurve variations at a given wavelength depending on the actual distribution of infrared absorbent in the atmosphere.

Studying the diversity of chemical regimes in the atmospheres of EChO targets is a necessary step to prepare one of the most exciting goals of XXI$^{st}$ century science: the search for signatures of life on extrasolar worlds. It is indeed crucial to understand all the types of abiotic (=without life involved) out-of-equilibrium processes and associated signatures that physics and chemistry can produce in planetary atmospheres before considering identifying those that life and only life can generate.

**Aknowledgements:**

O.V. acknowledges support from the KU Leuven IDO project IDO/10/2013 and from the FWO Postdoctoral Fellowship programme. M.A. acknowledges support from the European Research Council (ERC Grant 209622: E3ARTHs). Computer time for this study was provided by the computing facilities MCIA (Mésocentre de Calcul Intensif Aquitain) of the Université de Bordeaux and of the Université de Pau et des Pays de l'Adour.